\documentclass[10pt,oneside,english,letterpaper, reqno]{amsart}
\usepackage[T1]{fontenc}
\usepackage[latin1]{inputenc}
\usepackage{calc}
\usepackage{graphicx}
\usepackage{amssymb}
\IfFileExists{url.sty}{\usepackage{url}}
                      {\newcommand{\url}{\texttt}}

\makeatletter

\providecommand{\tabularnewline}{\\}

 \theoremstyle{plain}

\usepackage{latexsym}

\usepackage{cite}

\usepackage{babel}
\makeatother
\begin{document}


\title[Nadelstrahlung]{Lorenz fields convey energy as Nadelstrahlung}

\author{H. C. Potter}




\begin{abstract}
Gauge transformations leave unchanged only specific Maxwell fields.
To reveal more, I develop Lorenz field equations for superposed, sourced
and unsourced, wave function potentials. In this Maxwell form system,
the Lorenz condition is charge conservation. This allows me to define
three transformation classes that screen for Lorenz relevance. Nongauge,
sans gauge function, Lorentz conditions add polarization fields. These
enable emergent, light-like radiation. That from Lissajous potentials
is Nadelstrahlung. It conveys energy localized like particles at charge
conserving, progressive phase points. Such rays escape discovery in
modern Maxwell fields where gauge transformations suppress the polarizations.

\end{abstract}

\maketitle

\section{Introduction}

In 1867, during the time when J. C. Maxwell (1831-79) was publishing
his electromagnetic theory, L. V. Lorenz (1829-91) published his theory
equating light vibrations with electric currents \cite{Lorenz}. This
work is translated to modern vector notation and critiqued in \cite{Potter}.
Lorenz starts with Kirchhoff's Ohm's law expression. For scalar potentials
with a finite propagation speed, he obtains delayed potentials. These
potentials satisfy inhomogeneous wave equations and, when charge is
conserved, the eponic Lorenz condition \cite[pp. 268-9]{Whittaker}.
They are nonlocal. All their derivatives also must be inhomogeneous
wave function solutions for correspondingly differentiated remote
sources. Since the potentials satisfy wave equations they can be augmented
with homogeneous wave equation solutions. These sourceless augmentations
are local. Their derivatives depend only on proximate values. The
composite potentials still satisfy wave equations and the Lorenz condition
with one and the same propagation speed. But they are no longer strictly
delayed. Their differential character is mixed, local and nonlocal.

In his 1867 paper, Lorenz never develops first order field equations
from his potentials. In fact, until now, this has been ignored. With
magnetic induction as the vector potential curl, I obtain Lorenz field
equations with Maxwell form. This allows Lorenz delay to be fully
probed. For nonlocal potentials, the fields in these equations are
nonlocal also. The desirable locality can be restored as a far field
approximation for systems that satisfy specific size constraints \cite[p. 222]{Marion}.
For the augmented potentials, the Lorenz field equations incorporate
fields for electric displacement and magnetic field strength by adding
polarization fields. This development is presented formally in Sec.~\ref{sec:Lorenz-fields}.
It shows that wave function potentials satisfying the Lorenz condition
assure charge conservation. Thus, the form invariant Lorenz condition
is charge conservation. This limits potentials transformation. Polarization
radiation is described in Sec.~\ref{sec:Polarization-radiation}.
It should aid light mechanism study. Transformation and radiation
aspects are discussed in Sec.~\ref{sec:Discussion}.

\section{Lorenz fields\label{sec:Lorenz-fields}}

The electromagnetic Lorenz potentials satisfy differential wave equations
with charge, $\rho$, and current, $\mathbf{J}$, density sources
and, when charge is conserved, the Lorenz condition\begin{equation}
-a\boldsymbol{\nabla}\bullet\mathbf{A}=\frac{\partial\Omega}{\partial t}.\label{eq:delayedLorenzCondition}\end{equation}
 For sources distributed throughout all space, the potentials are
given by

\begin{subequations}\label{eqs:PotentialWaves} %
\begin{minipage}[b][0pt]{0.43\textwidth}%
\begin{equation}
\square_{a}\Omega=-4\pi\rho,\label{eq:OmegaWave}\end{equation}
\end{minipage}%
\quad{}\quad{}\hfill{}%
\begin{minipage}[b][0pt]{0.43\textwidth}%
 \begin{equation}
a\square_{a}\mathbf{A}=-4\pi\mathbf{\mathbf{J}}.\label{eq:Awave}\end{equation}
\end{minipage}%
\end{subequations}\hfill{}The wave equation operator for wave speed
$a$, $\square_{a}$, is defined in Sec.~\ref{sub:Potentials-transformation}.
When the potentials are augmented by solutions to the homogeneous
wave equation, $\square_{a}\Omega_{0}=\square_{a}\mathbf{A}_{0}=0$,
the Lorenz condition takes the modified form\begin{equation}
-a\boldsymbol{\nabla}\bullet(\mathbf{A}+\mathbf{A}_{0})=\frac{\partial(\Omega+\Omega_{0})}{\partial t}.\label{eq:AugmenteddelayedLorenzCondition}\end{equation}
 The augmentations could be set to zero by implicit scalar strengths.
For \\
 \hfill{}\begin{subequations}\label{eqs:EandB} %
\begin{minipage}[b][0pt]{0.45\textwidth}%
\begin{equation}
\mathbf{E}=-\boldsymbol{\nabla}\Omega-\mathring{\mathbf{A}}/a\label{eq:E}\end{equation}
\end{minipage}%
\quad{}and \hfill{}%
\begin{minipage}[b][0pt]{0.45\textwidth}%
 \begin{equation}
\mathbf{\mathbf{B}}=\mathbf{\boldsymbol{\nabla}}\times\mathbf{A},\label{eq:B}\end{equation}
\end{minipage}%
\hfill{}\end{subequations} \\
 we also have the fields $\mathbf{D}=\mathbf{E}+\mathbf{P}$ and $\mathbf{H}=\mathbf{B}-\mathbf{M}$.
The polarization fields are\\
 \hfill{}\begin{subequations}\label{eqs:PandM} %
\begin{minipage}[b][0pt]{0.45\textwidth}%
\begin{equation}
\mathbf{P}=-\nabla\Omega_{0}-\mathring{\mathbf{A}}{_{0}}/a\label{eq:P}\end{equation}
\end{minipage}%
\quad{}and \hfill{}%
\begin{minipage}[b][0pt]{0.45\textwidth}%
\begin{equation}
\mathbf{M}=-\boldsymbol{\nabla}\times\mathbf{A}_{0}.\label{eq:M}\end{equation}
\end{minipage}%
\hfill{}\end{subequations} \\
The ring denotes temporal partial differentiation.  Constants
for units adjustment are set to unit value. All fields satisfy wave
equations with propagation speed $a$. These definitions pass the
augmentation strengths to the polarizations. Without strengths, the
polarizations vanish when the augmentations are derived from a gauge
function, \emph{i.e.} when $\mathbf{A}_{0}=\boldsymbol{\nabla}\chi$
and $\Omega_{0}=-\mathring{\chi}/a$. The fields satisfy Maxwell form
equations:\\
 \hfill{}\begin{subequations}\label{eqs:FieldEquations} %
\begin{minipage}[b][0pt]{0.45\textwidth}%
\begin{equation}
\boldsymbol{\nabla}\bullet\mathbf{\mathbf{B}}=0,\label{eq:DivB}\end{equation}
\end{minipage}%
\quad{}\quad{}\hfill{}%
\begin{minipage}[b][0pt]{0.43\textwidth}%
 \begin{equation}
\boldsymbol{\nabla}\bullet\mathbf{D}=4\pi\rho,\label{eq:DivD}\end{equation}
\end{minipage}%

\begin{minipage}[b][0pt]{0.43\textwidth}%
\begin{equation}
a\mathbf{\boldsymbol{\nabla}}\times\mathbf{E}=-\partial\mathbf{\mathbf{B}}/\partial t,\label{eq:CurlE}\end{equation}
\end{minipage}%
\quad{}\quad{}\hfill{}%
\begin{minipage}[b][0pt]{0.43\textwidth}%
 \begin{equation}
a\mathbf{\boldsymbol{\nabla}}\times\mathbf{H}=\partial\mathbf{D}/\partial t+4\pi\mathbf{J}.\label{eq:CurlH}\end{equation}
\end{minipage}%
\hfill{}\end{subequations}Together Eqs.~(\ref{eq:DivD}) and (\ref{eq:CurlH})
give the condition for charge conservation\begin{equation}
\boldsymbol{\nabla}\bullet\mathbf{J}+\frac{\partial\rho}{\partial t}=0.\label{eq:ChargeContinuity}\end{equation}
 The Eq.~(\ref{eq:AugmenteddelayedLorenzCondition}) Lorenz condition
too gives this when acted on by the wave equation operator $\square_{a}$.

Fields derived from the delayed potentials can not be proportional
to those derived from the potential augmentations. This means coupling
normally provided by the constitutive relations, $\mathbf{D}=\epsilon\mathbf{E}$
and $\mathbf{B}=\mu\mathbf{H}$, is not primal. There are three choices.
First, take the relations as valid in very small space-time volumes
where boundary conditions are set. Second, further augment the potentials
with proportional delayed potentials. Then, wave function speed can
be adjusted for dielectric constant $\epsilon$ and magnetic permeability
$\mu$ by measurement. Third, 
couple the fields by local and nonlocal fluxes. To see this, notice
that the relation $\mathbf{H}\bullet(\boldsymbol{\nabla}\times\mathbf{D})-\mathbf{D}\bullet(\boldsymbol{\nabla}\times\mathbf{H})=\boldsymbol{\nabla}\bullet(\mathbf{D}\times\mathbf{H})$
allows two energy continuity expressions to be obtained from Eqs.~(\ref{eq:CurlE})
and (\ref{eq:CurlH}). When $\mathbf{E}$ and $\mathbf{B}$ are replaced
by $\mathbf{P}$ and $\mathbf{M}$ using the expressions for $\mathbf{D}$
and $\mathbf{H}$, \begin{subequations}\label{eqs:EnergyContinuity}
\begin{equation}
a\boldsymbol{\nabla}\bullet(\mathbf{H}\times\mathbf{\mathbf{D}})=\left[\mathbf{H}\bullet\frac{\partial\mathbf{H}}{\partial t}+\mathbf{D}\bullet\frac{\partial\mathbf{D}}{\partial t}\right]+4\pi\mathbf{\mathbf{D}}\bullet\mathbf{\mathbf{J}}.\label{eq:DxHcontinuity}\end{equation}
 More directly, \begin{equation}
a\boldsymbol{\nabla}\bullet(\mathbf{H}\times\mathbf{\mathbf{E}})=\left[\mathbf{H}\bullet\frac{\partial\mathbf{B}}{\partial t}+\mathbf{E}\bullet\frac{\partial\mathbf{D}}{\partial t}\right]+4\pi\mathbf{E}\bullet\mathbf{\mathbf{J}}.\label{eq:ExHcontinuity}\end{equation}
 \end{subequations} Both exist when the polarizations vanish. But
when $\mathbf{E}$ and $\mathbf{B}$ vanish, the latter vanishes and
the former carries the entire flux. This coupling allows local polarization
waves to emerge from a nonlocal electromagnetic field. The Hertz dipole
radiation solution discussed in Sec.~\ref{sub:Light-mechanization}
is the only extant example which exhibits this emergence.

\section{Polarization radiation\label{sec:Polarization-radiation}}

As sources go to zero the potentials become possible augmentations.
So, the potential augmentations can conserve charge. These augmentations
are then restrained by the Lorenz condition. Potentials in matter
free space have not been so restrained previously. With this restraint,
the potentials describe waves that can be considered to propagate
by progressively conserving charge. Features are given in Sec.~\ref{sub:Polarization-waves}.
There I show the polarizations will describe rays without the longitudinal
fields that plague other formulations \cite{Keller}. An example with
Lissajous vector potential is presented in Sec.~\ref{sub:Polarization-flux-localization}.
In this instance rays are real space paths determined by linearly
independent potential vector component phases. With neither wave packet
dispersion \cite{Darwin} nor quantum mechanical nonlocality, these
rays provide localization that has long eluded discovery. Localization
to rays is relaxed for linearly dependent component phases.

\subsection{Polarization waves\label{sub:Polarization-waves}}

As representative Lorenz potential augmentations consider

\begin{subequations}\label{eqs:AugPotentials} %
\begin{minipage}[b][0pt]{0.45\textwidth}%
\begin{equation}
\mathbf{A}_{0}=\left(f,g,h\right),\label{eq:Azero}\end{equation}
\end{minipage}%
\quad{}\hfill{}%
\begin{minipage}[b][0pt]{0.45\textwidth}%
 \begin{equation}
k\Omega_{0}=\mathbf{k}\bullet\mathbf{A}_{0}=k_{x}f+k_{y}g+k_{z}h.\label{eq:OmegaZero}\end{equation}
\end{minipage}%
\quad{}\hfill{}\end{subequations} When $f$, $g$ and $h$ are
functionally dependent only on phase factors $\omega t-\mathbf{k}\bullet\mathbf{r}$
where $\mathbf{k}=(k_{x},k_{y},k_{z})$ and $a^{2}k^{2}=\omega^{2}$,
these potentials satisfy the homogeneous wave equation and the Lorenz
condition. Since $\boldsymbol{\nabla}\Omega_{0}=-\mathring{\Omega}_{0}\mathbf{k}/\omega$
and $\mathbf{k}\times(\mathbf{k}\times\mathring{\mathbf{A}}_{0})=\mathbf{k}(\mathbf{k}\bullet\mathring{\mathbf{A}}_{0})-k^{2}\mathring{\mathbf{A}}_{0}$,

\begin{subequations}\label{eqs:Polarizations} %
\begin{minipage}[b][0pt]{0.45\textwidth}%
\begin{equation}
\omega{k}\mathbf{P}=\mathbf{k}\times(\mathbf{k}\times\mathring{\mathbf{A}}_{0}),\label{eq:wP}\end{equation}
\end{minipage}%
\quad{}\hfill{}%
\begin{minipage}[b][0pt]{0.45\textwidth}%
 \begin{equation}
{\omega}\mathbf{M}=\mathbf{k}\times\mathring{\mathbf{A}}_{0}.\label{eq:wM}\end{equation}
\end{minipage}%
\hfill{}\end{subequations}\\
 When the vector potential components are periodic functions they
can be considered to be plane waves. Unlike Lorenz fields defined
by nonlocal potentials, the polarizations are local functions. They
are orthogonal, $\mathbf{P}\bullet\mathbf{M}=0$, and have no longitudinal
components, $\mathbf{k}\bullet\mathbf{P}=\mathbf{k}\bullet\mathbf{M}=0$;
they carry a flux $\mathbf{M}\times\mathbf{P}=\omega^{-2}(\mathbf{k}\times\mathring{\mathbf{A}}_{0})^{2}\mathbf{k}/k$
equal to $\mathbf{D}\times\mathbf{H}$ when $\mathbf{E}=\mathbf{B}=0$
and travel indefinitely in a ray direction $\mathbf{k}$ at speed
$a$ without driving sources.

\subsection{Polarization flux localization\label{sub:Polarization-flux-localization}}

Polarization waves provide, at least, a classical solution to the
\cite{Duality1,Duality2,Rays,Piekara} duality. To see how, consider
a vector potential with Lissajous phases $\phi_{1}=\omega_{1}t-\mathbf{k}_{1}\bullet\mathbf{r}$,
$\phi_{2}=\omega_{2}t-\mathbf{k}_{2}\bullet\mathbf{r}$ and $\phi_{3}=\omega_{3}t-\mathbf{k}_{3}\bullet\mathbf{r}$
where $\mathbf{k}_{1}=(k_{1x},k_{1y},0)$, $\mathbf{k}_{2}=(k_{2x},k_{2y},0)$
and $\mathbf{k}_{3}=(0,0,k_{3z})$. Though periodic components are
not essential, added insight is provided by the simple form\begin{subequations}\label{eqs:ExPotentials}
\begin{equation}
\mathbf{A}_{0}=\left(\sin\phi_{1},\sin\phi_{2},\sin\phi_{3}\right).\label{eq:ExAzero}\end{equation}
 When $a^{2}k_{I}^{2}=\omega_{I}^{2}$ for $I=1,2\textrm{ or }3$,
$\square_{a}\mathbf{A}_{0}=0$. 
The Lorenz condition gives \begin{equation}
\Omega_{0}=(k_{1x}/k_{1})\sin\phi_{1}+(k_{2y}/k_{2})\sin\phi_{2}+\sin\phi_{3}.\label{eq:ExOmegaZero}\end{equation}
 \end{subequations}This also satisfies the homogeneous wave equation
$\square_{a}\Omega_{0}=0$. So these potentials can be taken to define
the Eqs.~(\ref{eqs:PandM}) Lorenz polarizations and flux given in
the Appendix. 

When $\phi_{1}=\phi_{2}=\phi$ for $\mathbf{k}_{1}\neq\mathbf{k}_{2}$
the phase moves on the parametric line \begin{subequations}\label{eq:ExShCoords}
\begin{equation}
X(t)=[k_{2y}(\omega_{1}t-\phi)-k_{1y}(\omega_{2}t-\phi)]/(k_{1x}k_{2y}-k_{2x}k_{1y}),\label{eq:ExShX}\end{equation}
 \begin{equation}
Y(t)=[-k_{2x}(\omega_{1}t-\phi)+k_{1x}(\omega_{2}t-\phi)]/(k_{1x}k_{2y}-k_{2x}k_{1y})\label{eq:ExShY}\end{equation}
 \end{subequations}with velocity\begin{equation}
\mathbf{V}=(k_{2y}\omega_{1}-k_{1y}\omega_{2},k_{1x}\omega_{2}-k_{2x}\omega_{1},0)/(k_{1x}k_{2y}-k_{2x}k_{1y})\label{eq:ExShV}\end{equation}
giving $V^{2}=2a^{2}k_{1}k_{2}[k_{1}k_{2}-(k_{1x}k_{2x}+k_{1y}k_{2y})]/(k_{1x}k_{2y}-k_{2x}k_{1y})^{2}$.
If the flux vector were to have a component perpendicular to the phase
velocity the ray would dissipate. So the vector product components
must vanish or, for the Eqs.~(\ref{eq:ExShFlux}) and (\ref{eq:ExShV})
flux and velocity,
\begin{equation}
(k_{2y}\omega_{1}-k_{1y}\omega_{2})(\frac{k_{2x}k_{2y}}{k_{2}}-\frac{k_{1y}^{2}}{k_{1}})-(k_{1x}\omega_{2}-k_{2x}\omega_{1})(-\frac{k_{1x}k_{1y}}{k_{1}}+\frac{k_{2x}^{2}}{k_{2}})=0.\label{eq:FxV}\end{equation}
 This reduces to 
$(k_{1y}+k_{2x})[k_{1}k_{2}-(k_{1x}k_{2x}+k_{2y}k_{1y})]=0$. 
So, $\mathbf{k}_{1}$ and $\mathbf{k}_{2}$ can satisfy two alternative
conditions for flux and phase velocity directions to be aligned. The
phase $\phi$ is a parameter that gives simple harmonic polarizations.
But $\phi$ is not a wave phase factor unless $\omega_{1}=\omega_{2}$,
because it is confined to the line defined by Eqs.~(\ref{eq:ExShCoords}).
As $\mathbf{k}_{1}\to\mathbf{k}_{2}$, the ray-like character persists.

\begin{figure}[htbp]
 \centering\scalebox{0.75}{\includegraphics{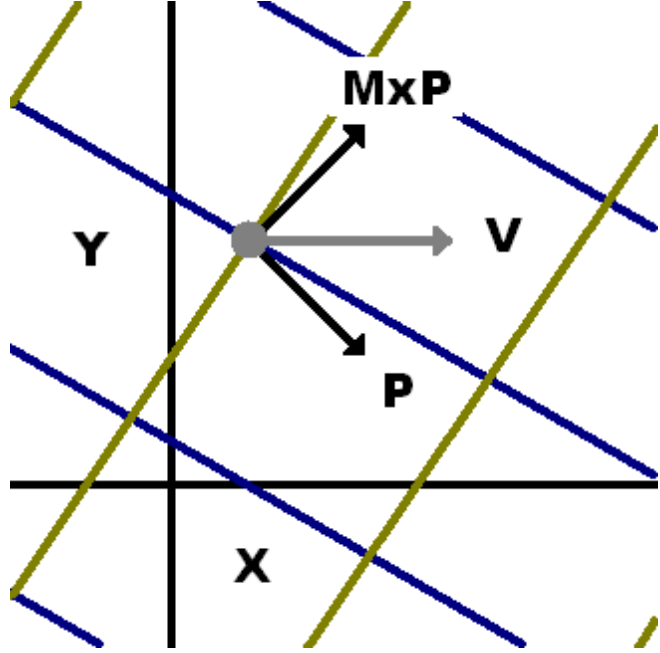}}

\caption{\label{fig:V_MxP}This figure depicts a phase and flux directions
non-alignment instance. Polarizations are defined by text Eqs.~(\ref{eqs:ExPotentials})
Lissajous potentials with vector potential component wave vectors
$\mathbf{k}_{1}=([(\sqrt{265}-3)/16]^{\frac{1}{2}},-\sqrt{3/8},0)$
and $\mathbf{k}_{2}=(1/\sqrt{2},\sqrt{3/2},0)$. For these potentials
the text Eq.~(\ref{eq:ExShV}) phase velocity, $\mathbf{V}$, equals
$(a,0,0)$. The hatching represents constant vector potential x- and
y-component values at phases $\phi_{1}=-7.60+2\pi n$ and $\phi_{2}=2.55+2\pi m$.
These travel with wave speed $a$ in their respective wave vector
directions. The Lissajous periodicities have the relative value $(\omega_{1}/\omega_{2})^{2}=(k_{1}/k_{2})^{2}=(\sqrt{265}+3)/32\approx0.60.$
The text Eq.~(\ref{eq:ExP}) electric polarization $\mathbf{P}$
at $(\phi_{1},\phi_{2})$ has components $\mathbf{P}_{1}$ normal
to $\mathbf{k}_{1}$ and $\mathbf{P}_{2}$ normal to $\mathbf{k}_{2}$.
These components depend independently on the $\mathbf{k}$'s, $\phi$'s
and vector potential component magnitudes to give a composite polarization
$\mathbf{P}$ and flux $\mathbf{M}\times\mathbf{P}$. Both $\mathbf{V}$
and $\mathbf{M}\times\mathbf{P}$ change magnitude and direction as
the $\mathbf{k}$'s are independently changed. But $\mathbf{M}\times\mathbf{P}$
can be brought into alignment with fixed $\mathbf{V}$ by changing
the vector potential component relative magnitudes and, equivalently
for the Eqs.~(\ref{eqs:ExPotentials}) potentials, the $\phi$'s.
This alignment is required to keep a ray from dissipating.}
\end{figure}

Although the case for $\mathbf{k}_{1}\neq\mathbf{k}_{2}$would appear
to have physical relevance in describing light-like waves, the more
general case in which $\phi_{1}\neq\phi_{2}$ may provide a particle
description for light. In this case, the phase point $(\phi_{1},\phi_{2})$
moves on a path defined by Eqs.~(\ref{eq:ExShCoords}) with $\phi$
replaced by $\phi_{1}$ or $\phi_{2}$. Now, the flux has the Eq.~(\ref{eq:ExFlux})
form. To prevent ray dissipation by flux components normal to the
phase path, restraints must again be applied to $\mathbf{k}_{1}$
and $\mathbf{k}_{2}$. These will depend on $\phi_{1}$ and $\phi_{2}$.
This means the flux is carried by phase points. This case is illustrated
in Fig.~\ref{fig:V_MxP} for select wave vector values that give
$\mathbf{V}=(a,0,0)$. Construction details are given in the caption.
Significantly, Fig.~\ref{fig:V_MxP} shows that $V$ can equal $a$
even when $\mathbf{V}$ and $\mathbf{M}\times\mathbf{P}$ are not
aligned. The concentrated energy bearers can
be marshaled into ensembles.%

\section{Discussion\label{sec:Discussion}}

Above, I have shown the Lorenz condition can select light-like rays
from electromagnetic fields. This permits light mechanism study. Also,
its equivalence to charge conservation limits potentials transformation.
These aspects are discussed below.

\subsection{Potentials transformation\label{sub:Potentials-transformation}}

Early last century, H. A. Lorentz (1853-1928) made two comments on
Maxwell field potentials. First, the Eqs.~(\ref{eqs:EandB}) electric
field strength and magnetic induction are unchanged in form for any
gauge function, $\chi$, that gives new potentials defined by\\
 \begin{subequations}\label{eq:GaugeTransform} %
\begin{minipage}[b][0pt]{0.45\textwidth}%
 \begin{equation}
\mathbf{A}=\mathbf{A}_{1}-\boldsymbol{\nabla}\chi\label{eq:A}\end{equation}
\end{minipage}%
\quad{}and \hfill{}%
\begin{minipage}[b][0pt]{0.45\textwidth}%
 \begin{equation}
\Omega=\Omega_{1}+\mathring{\chi}/a.\label{eq:Omega}\end{equation}
\end{minipage}%
\hfill{}\end{subequations} Gauge restraints are now widely accepted
\cite{JacksonOkun}. Some gauges do not even specify a gauge function.
For these, a gauge function further restrains the potentials. 
The second comment was that Eq.~(\ref{eq:delayedLorenzCondition}),
dubbed the Lorentz condition, causes the Maxwell field potentials
to satisfy the Eqs.~(\ref{eqs:PotentialWaves}) wave equations. This
is shown in \cite{Heras}. Lorentz did not relate these comments.
So, the Lorentz condition has been treated as a gauge that can be
freely changed. This freedom and wave function potential nonlocality
have caused potential reality to be denied. When Eq.~(\ref{eq:delayedLorenzCondition})
is charge conservation, as I have shown above that it is, these comments
are coupled. As charge conservation, Eq.~(\ref{eq:delayedLorenzCondition})
is a basic law. With this, in Sec.~\ref{sec:Lorenz-fields} I have
shown the Eqs.~(\ref{eqs:PotentialWaves}) wave function potentials
give the Eq.~(\ref{eqs:FieldEquations}) field equations with Maxwell
form. This is the second comment converse. Thus, Maxwell fields must
always have wave function potentials. The potentials can be local,
nonlocal or mixed. Further, as a basic law, Eq.~(\ref{eq:delayedLorenzCondition})
can not be freely changed.

When the original potentials are sourced wave functions Eqs.~(\ref{eq:GaugeTransform})
resolve to three Lorenz classes that need not have the gauge function
form. First, the new potentials are not wave functions. Second, the
new potentials are wave functions with changed source values. Third,
the new potentials are wave functions with unchanged source values.
In the first case the Lorentz condition will not apply. Then, charge
will not be conserved and the potentials will not propagate with a
wave speed. Both must be spurned. The latter, because tests in \cite{delay1,delay2,delay3,delay4}
show longitudinal electric fields to propagate with finite speed.
The last two cases will conserve charge if the Lorentz condition persists.
This persistence will assure Lorenz field equations in which charge
and current density are deemed observables that produce potentials
and fields yielding emergent light-like waves. These conserve charge
progressively as described in Sec.~\ref{sub:Polarization-waves}.
Further, localized energy is conveyed on ray forming phase points
as described in Sec.~\ref{sub:Polarization-flux-localization}. For
gauge transformations, persistence requires the Lorenz gauge, $\square_{a}\chi\equiv\boldsymbol{\nabla}\bullet\boldsymbol{\nabla}\chi-\overset{\circ\circ}{\chi}/a^{2}=0$.
It suppresses the polarizations and, thus, their dependent, emergent
rays.

\subsection{Light mechanization\label{sub:Light-mechanization}}

In the late eighteen hundreds, H. von Helmholtz (1821-94) tried to
unite electromagnetic theories \cite{Woodruff}. To this end he based
his concept on electric and magnetic polarization. From this he obtained
wave equations for a homogeneous medium. His transverse waves have
speed $c/\sqrt{\epsilon\mu}$. Here $c$ is the vacuum light speed.
Only difficult, precision measurements could complete the theory,
because his longitudinal wave speed has any value greater than or
equal to zero. Unlike Helmholtz waves, Lorenz waves are simply transverse,
because Eq.~(\ref{eq:DivD}) gives $\boldsymbol{\nabla}\bullet\mathbf{P}=0$.

To promote his concept, Helmholtz issued a challenge to measure a
coupling between electromagnetism and dielectric polarization. His
former student, H. Hertz (1857-94), later claimed the prize. He then
went on to observe dipole radiation reflection and interference. Based
on this Hertz concluded that polarization propagation is like vacuum
light \cite[pp. 19 and 122-3]{Hertz} and \cite{Hertz2}. Hertz's
dipole field \cite{Marion,Hertz,Hertz2} has radial, transverse wave
emergence in the far field. At intermediate distances the waves have
greater than light speed that approaches light speed in the far field.
These waves change from longitudinal to transverse as the radial direction
changes from dipole length to dipole equator. Hertz reconciled these
properties in the equatorial plane by observing strobed interference
between free air and straight wire waves \cite[pp. 150-5]{Hertz}.
His and recent \cite{PhaseSkip} reports suggest light structure may
be open to study. Antiphased fields in \cite{PhaseSkip} may represent
an Eq.~(\ref{eq:CurlH}) Ampere law based, magnetoinductive internal
structure \cite{Potter}.

For vacuum polarization waves Eq.~(\ref{eq:DxHcontinuity}) takes
a simple form. 
By the Gauss theorem, it represents an equality between the temporal
energy change in a volume and the energy flux through its surface.
For monochromatic waves, $\mathbf{M}\times\mathbf{P}$ can be written
using the Hertz analogy as $h\nu\mathbf{f}$ where $h\nu$ is photon
energy and $\mathbf{f}$ is photon flux with coherence length inversely
related to monochromaticity departure \cite{Mandel}. This dependence
means that the photon power density must approach zero as the coherence
length becomes very large for monochromatic photons. The compact,
quantum particle photon concept is untenable in this limit. The concept
is further threaten by confounding photon size with wavelength. This
problem is revealed by the Table \ref{cap:Some-Vacuum-Photon} benchmarks.
There the lowest energy photons have a wavelength greater than the
Earth orbit radius. These conflicts are homologized in \cite{Piekara}
by setting a hypothetical photon energy density equal to a flux magnitude
divided by the energy transfer speed.%
\begin{table}[h]

\caption{Some Vacuum Photon Properties\label{cap:Some-Vacuum-Photon}}

\begin{centering}\begin{tabular}{lll|lll}
\hline 
Energy&
Frequency&
Wavelength&
Energy&
Frequency&
Wavelength\tabularnewline
{\quad{}{\small {$h\nu$}}}&
{\quad{}{\small {$\nu$ Hz}}}&
{\quad{}{\small {$\lambda$ cm}}}&
{\quad{}{\small {$h\nu$}}}&
{\quad{}{\small {$\nu$ Hz}}}&
{\quad{}{\small {$\lambda$ cm}}}\tabularnewline
\hline 
1 GeV &
0.241E24&
12.4E-14&
1 $\mu$eV \cite{Hertz}&
0.241E9&
12.4E1\tabularnewline
1 MeV &
0.241E21&
12.4E-11&
1 neV&
0.241E6&
12.4E4\tabularnewline
1 keV &
0.241E18&
12.4E-8&
1 peV &
0.241E3&
12.4E7\tabularnewline
1 eV &
0.241E15&
12.4E-5&
1 feV&
0.241&
12.4E10\tabularnewline
1 meV&
0.241E12&
12.4E-2&
1 aeV \cite{PhaseSkip}&
0.241E-3&
12.4E13\tabularnewline
\hline
\end{tabular}\par\end{centering}

\noindent \begin{raggedright}{\small For energy, $h$ is Planck's
constant. Electron rest mass equals 0.5 MeV, visible light at 5000
$\mathring{\textrm{A}}$ equals 2.48 eV and cosmic microwave background
radiation at 3 K equals 0.26 meV. Atomic nucleus radii are about E-12
cm, atomic radii are about E-8 cm, Earth radius is 6.38E8 cm and Earth
orbit radius is 1.5E13 cm. Solar radiant energy flux at Earth is 8.58E21
eV/(s m$^{2}$) with an energy distribution that should be appropriate
for the 5780 K effective sun temperature. The} H. Hertz, \emph{Electric
Waves}, D. E. Jones, trans. (Dover Pubs., Inc., NY, 1962) {\small and}
P. J. Chi and C. T. Russell, {}``Phase skipping and Poynting flux
of continuous pulsations,'' J. Geophys. Res. $\mathbf{103}$:A12,
29479-29491(1998) {\small citations attempt internal structure study. }\par\end{raggedright}
\end{table}

When phase point motion and wave flux are not aligned, flux normal
to the phase point motion would force ray dissipation. This is like
saying that all light rays are electromagnetic, but not all electromagnetic
waves are light rays. Failure to recognize this has prevented light-like
rays from being found in electromagnetic fields to help describe light
behavior in optics and photography \cite{Duality1,Duality2,Rays}.
Synchrotron light emitted as rays by high speed electrons in a circular
orbit supports this. These rays are attributed to the radial acceleration
not the periodic linear acceleration required to maintain the electron
energy. Their presence in the electric field has not been shown \cite{Synchrotron1,Synchrotron2,Synchrotron3}.
However, one unconfirmed report \cite{Elder} finds them to be electrically
polarized in the electron orbit plane like the Sec.~\ref{sub:Polarization-flux-localization}
phase directed polarization waves.

Although phase independence is compelling, its nonexistence would
support treating rays as having the closely bound potentials described
in Sec.~\ref{sub:Polarization-waves} above. In this limit energy
should still be borne progressively by charge conserving phase points.
Even so, phase independence is a hidden variable whose consequence
is unintuitive. Whether electromagnetic energy localization to phase
points is consistent with quantum statistics should be examined elsewhere,
because it has special features to help model the standard quantum
oscillators. The points are defined by Lissajous potentials. They
bear a fixed flux at constant speed. They vanish where either charge
cannot be conserved or flux and phase are not aligned. Individually,
they do not bear oscillations. But they can be marshaled into Eq.~(\ref{eq:ExSmShFlux^{2}}),
Piekara bead-chain, flux magnitude beads. These features provide a
new means to help describe light generation. If flux and phase alignment
were required for light generation, its low intrinsic likelihood would
predict a small efficiency. One challenge is light from atoms. Its
study could help increase laser output when atoms are aligned by shell
structure.

\section{Conclusion}

I have shown that if Lorenz had published field equations electromagnetism
as we know it today would have a solid, delay based etiology. We would
have Lorenz condition equivalence to charge conservation, light-like
ray emergence from fields and energy conveyance by field phase points.
Based on this I propose that the terms {}``Lorenz condition\char`\"{}
and {}``Lorentz condition\char`\"{} be retained. The Lorentz condition
would be only a potentials transformation. The Lorenz condition would
express charge conservation in a form left unchanged by potentials
transformation, Lorenz covariance. Further, the light-like rays that
convey localized energy in Lorenz fields escape discovery in modern
Maxwell fields. So, they may aid photon mechanization. 

\appendix

\section*{Appendix}

For the Eqs.~(\ref{eqs:ExPotentials}) potentials, Eqs.~(\ref{eqs:PandM})
gives \begin{subequations}\label{eqs:Ex}\begin{align}
\mathbf{P} & =(\frac{k_{2x}k_{2y}}{k_{2}}\cos\phi_{2}-\frac{k_{1y}^{2}}{k_{1}}\cos\phi_{1},\frac{k_{1x}k_{1y}}{k_{1}}\cos\phi_{1}-\frac{k_{2x}^{2}}{k_{2}}\cos\phi_{2},0),\label{eq:ExP}\\
\mathbf{M} & =(0,0,k_{2x}\cos\phi_{2}-k_{1y}\cos\phi_{1})\quad\textrm{and}\label{eq:ExM}\end{align}
 \begin{equation}
\begin{split}\mathbf{M}\times\mathbf{P} & =(k_{2x}\cos\phi_{2}-k_{1y}\cos\phi_{1})\times\\
 & \times(-\frac{k_{1x}k_{1y}}{k_{1}}\cos\phi_{1}+\frac{k_{2x}^{2}}{k_{2}}\cos\phi_{2},\frac{k_{2x}k_{2y}}{k_{2}}\cos\phi_{2}-\frac{k_{1y}^{2}}{k_{1}}\cos\phi_{1},0).\end{split}
\label{eq:ExFlux}\end{equation}
\end{subequations} Taking $\phi_{1}=\phi_{2}=\phi$ gives the simple
harmonic polarizations and flux

\begin{subequations}\label{eqs:ExSh} \begin{align}
\mathbf{P} & =\cos\phi(\frac{k_{2x}k_{2y}}{k_{2}}-\frac{k_{1y}^{2}}{k_{1}},\frac{k_{1x}k_{1y}}{k_{1}}-\frac{k_{2x}^{2}}{k_{2}},0),\label{eq:ExShP}\\
\mathbf{M} & =\cos\phi(0,0,k_{2x}-k_{1y})\quad\textrm{and}\label{eq:ExShM}\\
\mathbf{M}\times\mathbf{P} & =(k_{2x}-k_{1y})\cos^{2}\phi(-\frac{k_{1x}k_{1y}}{k_{1}}+\frac{k_{2x}^{2}}{k_{2}},\frac{k_{2x}k_{2y}}{k_{2}}-\frac{k_{1y}^{2}}{k_{1}},0)\label{eq:ExShFlux}\end{align}
 \end{subequations} having bead-chain \cite{Piekara} flux squared
magnitude\begin{equation}
\left|\mathbf{M}\times\mathbf{P}\right|^{2}=(k_{2x}-k_{1y})^{2}[k_{2x}^{2}-2\frac{k_{2x}k_{1y}}{k_{1}k_{2}}(k_{2x}k_{1x}+k_{2y}k_{1y})+k_{1y}^{2}]\cos^{4}\phi.\label{eq:ExShFlux^{2}}\end{equation}

When $\mathbf{k}_{1}=\mathbf{k}_{2}=\mathbf{k}=(k_{x},k_{y},0)$,
the polarizations further simplify to \begin{subequations}\label{eqs:ExSmSh}
\begin{align}
k\mathbf{P} & =(k_{x}-k_{y})\cos\phi(k_{y},-k_{x},0)\quad\textrm{and}\label{eq:ExSmShP}\\
\mathbf{M} & =\cos\phi(0,0,k_{x}-k_{y})\label{eq:ExSmShM}\end{align}
 \end{subequations} with the flux having direction $\mathbf{k}$
and magnitude\begin{equation}
\left|\mathbf{M}\times\mathbf{P}\right|=(k_{x}-k_{y})^{2}\cos^{2}\phi.\label{eq:ExSmShFlux^{2}}\end{equation}
 This flux is highly anisotropic with maximum values for $k_{x}=-k_{y}$
and null values for $k_{x}=k_{y}$.

\renewcommand{\refname}{\normalsize References and links \rm}

\end{document}